\documentclass[twocolumn]{aastex631}

\usepackage{booktabs}
\usepackage{amsmath}
\usepackage{xspace}
\newcommand{\unit}[2]{\ensuremath{\textrm{#1}^{#2}}}

\newcommand{\Msun}{\ensuremath{M_{\odot}}}

\newcommand{\cosmogems}{\textsf{CosmoGEMS}\xspace}
\newcommand{\CMC}{\textsf{CMC}\xspace}

\begin{document}

%\title{Toward Realistic Globular Cluster Stream Models in a Fully Cosmological Context}
\title{Breaking Down the \cosmogems: Toward Modeling and Understanding Globular Cluster Stellar Streams in a Fully Cosmological Context}

%\correspondingauthor{}
%\email{}

\author[0000-0001-5214-8822]{Nondh Panithanpaisal}
\affiliation{\carnegie}
\affiliation{\tapir}

\author[0000-0003-3939-3297]{Robyn E. Sanderson}
\affiliation{\upenn}
%\affiliation{\cca}

\author[0000-0003-4175-8881]{Carl L. Rodriguez}
\affiliation{\unc}

\author[0000-0003-2539-8206]{Tjitske Starkenburg}
\affiliation{\ciera}
\affiliation{\northwestern}
\affiliation{\skai}

\author[0000-0003-0256-5446]{Sarah~Pearson}
\affiliation{\affNBI}

\author[0000-0002-7846-9787]{Ana Bonaca}
\affiliation{\carnegie}

\author[0000-0003-3729-1684]{Philip F. Hopkins}
\affiliation{\tapir}

\author[0000-0003-0341-6928]{Brian T.~Cook}
\affiliation{\unc}

\author[0000-0002-8354-7356]{Arpit Arora}
\affiliation{Department of Astronomy, University of Washington, Seattle, WA 98195, USA}

\author[0000-0002-9660-9085]{Newlin C. Weatherford}
\affiliation{\carnegie}

%\author{Please add your affiliations or send them to me}

% affiliations 
\newcommand{\upenn}{Department of Physics \& Astronomy, University of Pennsylvania, 209 S 33rd St, Philadelphia, PA 19104, USA}
\newcommand{\cca}{Center for Computational Astrophysics, Flatiron Institute, 162 5th Ave, New York, NY 10010, USA}
\newcommand{\carnegie}{The Observatories of the Carnegie Institution for Science, 813 Santa Barbara St, Pasadena, CA 91101, USA}
\newcommand{\tapir}{TAPIR, Mailcode 350-17, California Institute of Technology, Pasadena, CA 91125, USA}
\newcommand{\affNBI}{DARK, Niels Bohr Institute, University of Copenhagen, Jagtvej 155A, 2200 Copenhagen, Denmark}
\newcommand{\unc}{Department of Physics and Astronomy, University of North Carolina at Chapel Hill, 120 E. Cameron Ave, Chapel Hill, NC, 27599, USA}
\newcommand{\ciera}{Center for Interdisciplinary Exploration and Research in Astrophysics (CIERA), Northwestern University, 1800 Sherman Ave, Evanston IL 60201, USA}
\newcommand{\northwestern}{Department of Physics and Astronomy, Northwestern University, 2145 Sheridan Rd, Evanston IL 60208, USA}
\newcommand{\skai}{NSF-Simons AI Institute for the Sky (SkAI), 172 E. Chestnut St., Chicago, IL 60611, USA}

\begin{abstract} 
Next-generation surveys are expected to uncover thousands of globular cluster (GC) stellar streams, motivating the need for a theoretical framework that produces realistic GC streams in a fully cosmological, Milky Way-like environment.
We present \cosmogems, a star-by-star cosmological GC stream framework that self-consistently links small-scale cluster physics with large-scale Galactic dynamics. The initial phase-space positions of stream stars are informed by post-processed GC populations within the FIRE cosmological simulation. Escaped stars are orbit-integrated from their time of escape to the present day in a time-evolving Galactic potential extracted from the same simulation using a basis function expansion.
We explore two example streams on different orbits. One forms a long, thin stream with a velocity dispersion consistent with Milky Way GC streams. However, it exhibits a clump and orbital-phase-dependent misalignments due to the evolving potential. The other stream develops both a thin component and a diffuse, shell-like structure, similar to features observed in streams like Jhelum.
These results highlight the power of fully cosmological models in producing realistic stream morphologies and kinematics. Unlike idealized simulations, our models naturally incorporate time-dependent changes in the progenitor’s orbit, including orbital plane evolution, which significantly affects stream structure. This challenges common assumptions in stream-finding algorithms and interpretation.
\cosmogems provides a key step toward connecting future stellar stream observations with the physics of globular cluster evolution and hierarchical galaxy formation in a cosmological context.
\end{abstract}

\section{Introduction} \label{sec:intro}

%Mass segragation in GCs and their mass loss over time. This affect streams (Balbinot+2018, and more), but previous focus has mostly been on mass functions of streams compared to their parent globular clusters (Balbinot+2018) and the effect of black hole formation and growth in clusters (Roberts+2024). 

There are over one hundred known stellar streams around our own Milky Way (MW). Of these, $\sim80$ streams are thought to originate from globular clusters (GCs); \citep{mateu2023,bonaca2025}. This number is expected to grow significantly in the near future, as next-generation surveys --- such as Euclid, the Rubin Observatory LSST, and the Nancy Grace Roman Space Telescope \citep{euclid,lsst,roman} --- will be capable of detecting not only fainter stellar streams in the MW halo \citep{pearson2024} but also extragalactic GC stellar streams and their density variations for the first time \citep{pearson2019, pearson2022, aganze2024}. These tidally disrupted GCs produce long, cold stellar streams that are valuable for constraining the properties of their host galaxies \citep[e.g.,][]{reino2021, palau2023, ibata2024, nibauer2025b} and dark matter subhalos \citep[e.g.,][]{johnston2002, ibata2002, yoon2011, carlberg2012, carlberg2013,price-whelan2018,carlberg2020,banik2021}. 

While it is now possible to study GC populations using post-processing techniques applied directly to the output of cosmological simulations \citep{chen2022, chen2023, gbof1, gbof2}, attempts to predict the observable population of GC streams in the MW have thus far been limited to static potentials and simplified stream modeling \citep{pearson2024}. As a result, establing a comprehensive theoretical model that self-consistently accounts for GC formation and disruption, while also tracking the orbits of escaped stars, remains an unresolved challenge. The absence of such a self-consistent theoretical framework significantly hinders our ability to predict and interpret future observations, especially in the detailed stream morphology.

%There is currently no theoretical framework that produces realistic GC stream populations in the MW or nearby galaxies, severely hindering our ability to predict and interpret future observations. Without such a theoretical framework, even simple questions --- such as how many GC streams exist around MW-mass galaxies and how many do we expect to observe --- are extremely difficult to answer.

Due to its low progenitor mass and the dense, collisional nature of globular clusters, simulating the formation of a GC stream within cosmologically evolving hosts is a computationally challenging task. It requires resolving both the small-scale collisional physics of the cluster and the large-scale evolution of the galactic tidal field. Specifically, the mechanisms by which stars escape from the cluster are driven by several physical processes \citep[e.g., see discussions in][]{weatherford2023,weatherford2024}. Even in complete isolation, a cluster loses mass through two-body relaxation --- a process in which stars gradually diffuse to positive total energy via many weak, uncorrelated encounters \citep{gnedin1997, vesperini1997, baumgardt2003}. This process is further enhanced in the presence of a background galactic potential, which effectively lowers the energy threshold for escape. In addition to weak encounters, stars can also be ejected through strong gravitational scattering, involving close two-body or three-body interactions \citep{cabrera2023,weatherford2023}. Finally, many stars escape due to global Galactic tidal effects, such as tidal shocking when the cluster passes through the disk \citep[e.g.,][]{spitzer1987book}.

Traditional efforts to model GC streams often bypass modeling the internal dynamics of the cluster entirely, instead using ``particle spray'' models, where tracer particles are released from the cluster’s Lagrange points with various prescriptions for their initial phase-space distributions and release times \citep{varghese2011,kupper2012,lane2012,bonaca2014,gibbons2014,fardal2015}. These models are computationally efficient and can accurately reproduce the overall stream tracks, leading to valuable constraints on the shape and mass distribution of the MW's halo \citep[e.g.,][]{gibbons2014, kupper2015, pearson2015, bovy2016}. 

However, particle spray models typically rely on numerous simplifying assumptions, making certain stream features, such as density variations, more difficult to interpret. For example, these models often release massless tracer particles at fixed time intervals, with their dynamical properties sampled from a prescribed phase-space distribution. In addition, they require assumptions about the Galactic potential and, in some cases, the self-gravity of the cluster—both of which evolve over time and may be poorly constrained. Since the full time evolution of the MW mass distribution is not known, the Galactic potential is often assumed to be analytical and static, typically consisting of three or more components representing standard galactic structures such as the bulge, disk, and halo \citep{bovy2015, mcmillan2017, gala}.

Finally, the stream progenitor is generally assumed to follow some initial orbit, especially since many known GC streams no longer have observable progenitors. While some existing GC stream models have attempted to relax or improve upon some of the individual assumptions \citep[e.g.,][]{pearson2024,chen2025} of the particle spray model, none has yet addressed all of them in a fully self-consistent way.

For the first time, we present \cosmogems (Cosmological Globular cluster streams Exploration and Modeling with Simulations), an integrated post-processing approach for producing a GC stream in a cosmologically evolving host galaxy potential. \cosmogems self-consistently bridges the cluster evolution physics with the galaxy formation physics, producing GC streams at a star-by-star resolution. We follow the orbit of individual stripped stars and show the effect of internal globular cluster evolution on the formation and stellar content of its stellar stream. We show proof-of-concept with two example cosmological GC streams formed with our pipeline.

%\FIXME{Double check that the intro is not too wordy.}

%In this letter, we present an integrated approach to stream formation from globular clusters in a cosmologically evolving host galaxy potential. We show the effect of internal globular cluster evolution on the formation and stellar content of its stellar stream. We follow the orbit of individual stripped stars and quantify the present-day observability of the stellar stream using current and future facilities based on it's stellar population. We discuss the influence of the cluster orbit on the mass segregation and observability of the stellar stream.

\section{Methods} \label{sec:methods}

\begin{figure}
\plotone{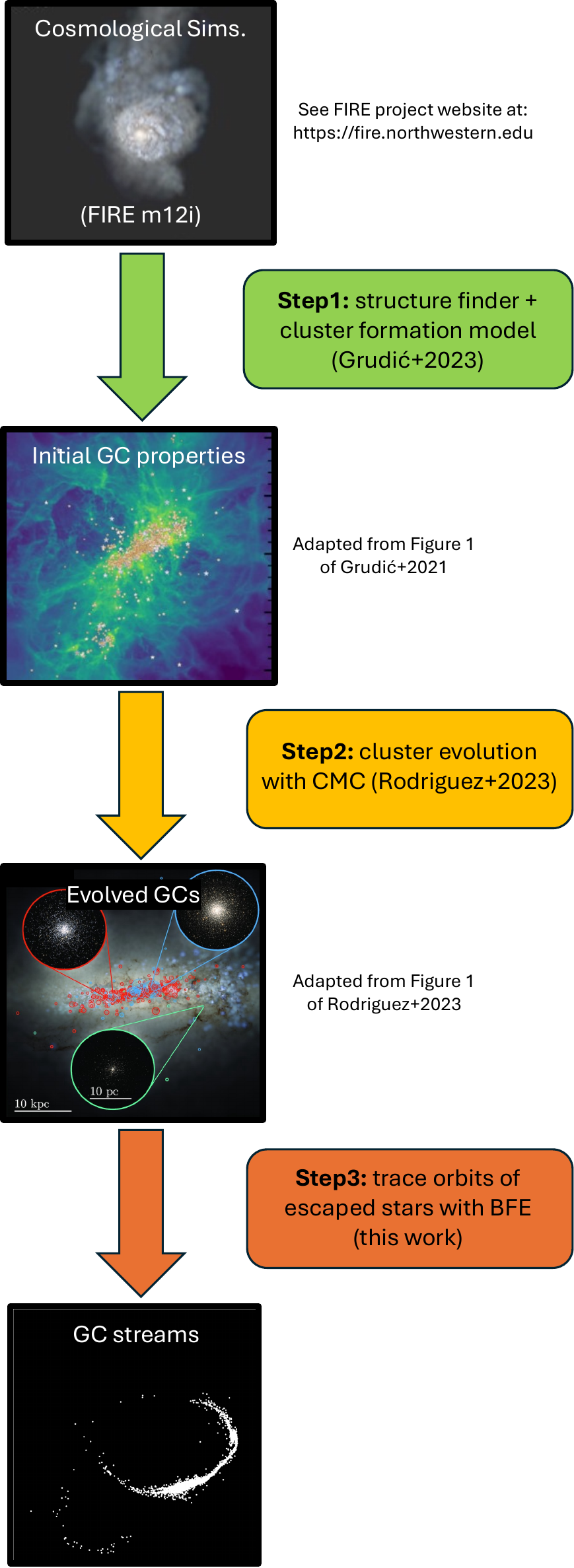}
\caption{A diagram describing steps in our cosmological GC stream production pipeline, \cosmogems. In step 1, the initial cluster properties were derived from applying structure finder and cluster formation model on the FIRE \textsf{m12i} simulation snapshots \citep{gbof1}. Next, these initial clusters were evolved to $z=0$ using the collisional \CMC model \citep{gbof2} in step 2. Finally, in step 3, we trace orbits of individual escaped stars within the Galaxy. \label{fig:gc_production_pipeline}}
\end{figure}

In this section, we present our star-by-star cosmological GC stream generation, \cosmogems, as shown in Figure \ref{fig:gc_production_pipeline}. The process involves three main steps, each of which is described in detail below. These include the cosmological simulation, the cluster formation and evolution model, and the orbit integration of the escaped stars.

\subsection{FIRE-2 simulation} \label{subsec:sim}
Our framework is based on the halo model \textsf{m12i} that was simulated as part of the FIRE-2 simulation suite of zoom-in baryonic simulations of MW-mass galaxy formation \citep{wetzel2016,hopkins2018,wetzel2023}. This simulation was run with magnetohydrodynamic version of the \textsf{GIZMO} code \citep{hopkins2015}. We also chose the version that includes prescriptions for conduction and viscosity for a more physically complete picture. However, we note that the effects of these added processes are not significantly changing the evolution and the properties of the host galaxy at the present day \citep{su2017,hopkins2018}. The overall host galaxy properties at $z=0$ are comparable to the standard metal diffusion run first presented in \cite{wetzel2016}. At the present day, the host galaxy has stellar mass of $M_\star=6.7\times10^{10}\Msun$ and halo mass of $M_{200m}=1.2\times10^{12}\Msun$, broadly consistent with the mass estimates for the MW \citep{BlandHawthorn2016}.

Each star particle in the simulation ($m_b\sim7000\Msun$) represents a single stellar population with the same ages and metallicities. Hence, GCs with typical mass of $M_\star\sim10^4\Msun$ do not natively form. To study the GC population, post-processing techniques are required.

%The cluster was evolved using the Cluster Monte Carlo (CMC) code, while the galactic potentials were fitted to individual cosmological simulation snapshots from FIRE simulation of MW-mass galaxy using a set of basis functions. In this framework, each progenitor orbit is represented by a tracer particle orbit extracted directly from the simulation snapshots.

\subsection{Step 1 \& 2: globular cluster formation and evolution}
\label{subsec:gcformation}
The cluster formation and evolution model that we used were described in full in \cite{gbof1} and \cite{gbof2}, hereafter Great Balls of FIRE I (GBoF1) and Great Balls of FIRE II (GBoF2), which we briefly summarize here.

Due to the resolution, the simulation resolves the bulk properties of giant molecular clouds (GMCs) of mass $\gtrsim10^5\Msun$ \citep{guszejnov2020}. A cluster formation model, calibrated to the much higher resolution simulations \citep{grudic2021sim}, was then applied to accurately determine properties of the clusters formed inside these GMCs. This is a statistical model that depends on the mass, size, and metallicity of the GMC (see Section 2.3 of GBoF1 for more details). We note that the model does not reproduce the age and metallicity statistics of old MW GCs ($\gtrsim11$ Gyr old), possibly due to the difference in the early star formation histories between the MW and \textsf{m12i}(see Figure 13 in GBoF1). 

In GBoF2, 895 of these clusters were then evolved up to $z=0$ using the Cluster Monte Carlo \citep[\CMC][]{joshi2000,pattabiraman2013,cmc} code. CMC allows for a time-varying tidal field, specified by a tidal tensor extracted directly from the same simulated galaxy \textsf{m12i}. Unlike the traditional N-body code where all pairwise forces are summed over, \CMC assumes each cluster to be spherical and treats many two-body encounters as a single effective encounter, i.e., the H{\'e}non's method \citep{henon1971}. However, comparing to the traditional N-body integrator NBODY6 \citep{webb2014}, \CMC has been shown to predict a reliable cluster's mass-loss rate \citep{gbof2}. To extract the host galaxy tidal forces, GBoF2 first identified a single star particle that was associated with the cluster's parent GMC. This tracer particle was then tracked across all simulation snapshots up to $z=0$. The cluster stars were initialized with masses sampled from the \cite{kroupa2001} initial mass function between 0.08 and 150 $\Msun$. Moreover, 10\% of them were randomly chosen to be in binary systems with their mass ratios drawn from a uniform distribution between 0.1 and 1, typical for N-body simulations of star clusters. \CMC is coupled to the \textsf{COSMIC} code for stellar population synthesis \citep{Breivik2020}.  \textsf{COSMIC} evolves stars using fitting formula for stellar evolution, and models binary interactions using detailed prescriptions (tides, mass transfer, supernova kicks).  This allows us to model stellar and binary evolution in \CMC self-consistently from the zero-age main sequence to the present day.

In GBoF2, once a star was ``ejected'' from the cluster based on their radial stripping criteria (see their Appendix A), it was instantaneously removed from the simulation, since their goal was to study the present-day GC population. However, in this work we focus on the evolution of the escaped stars within the host galaxy --- \textit{where do the escaped stars go once they leave the cluster?}

\begin{figure}
\plotone{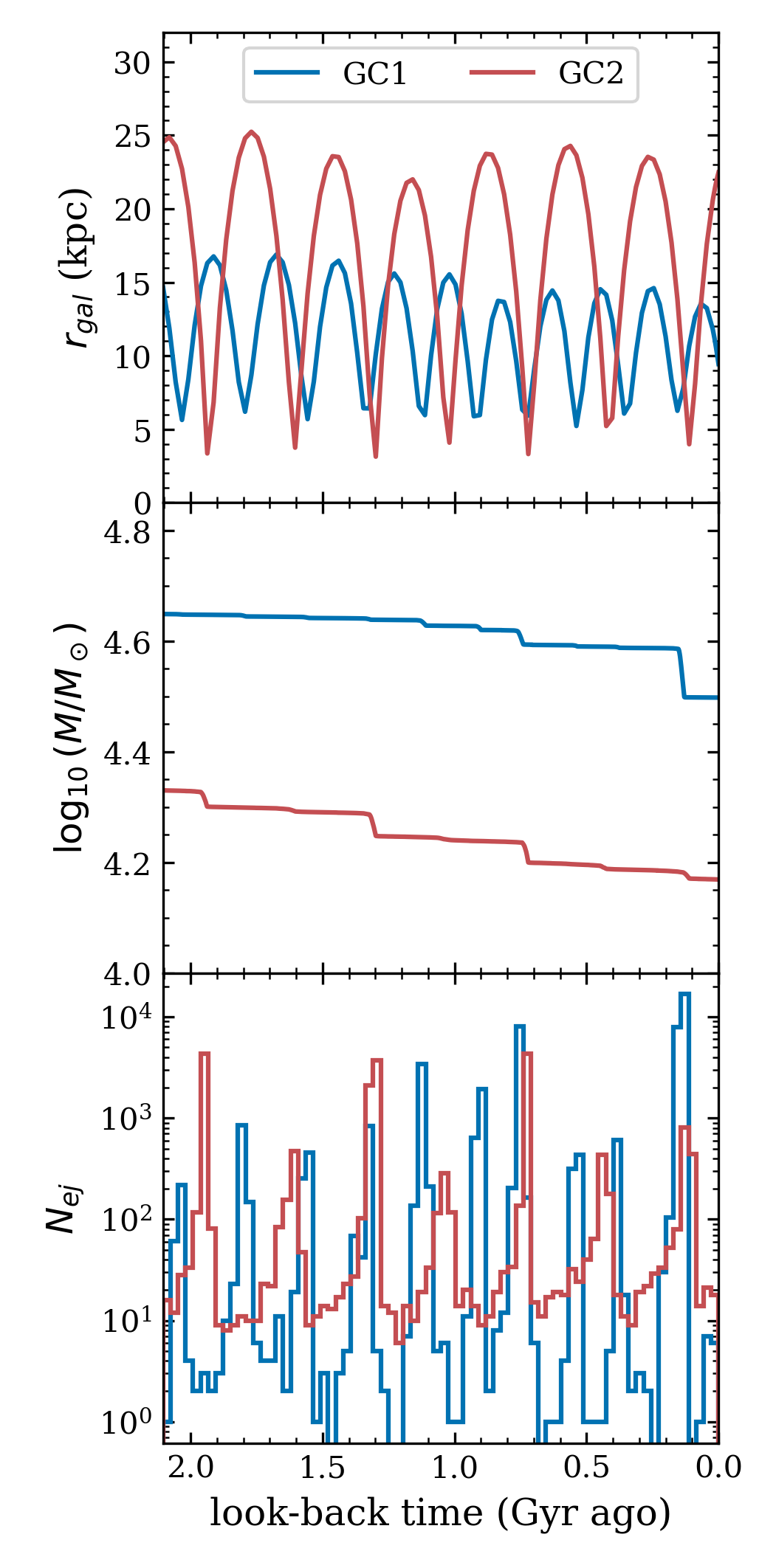}
\caption{Properties of GC1 and GC2 as a function of look-back time in blue and red, respectively. The top panel shows the the 3D galactocentric (physical) distances of the clusters, $r_{gal}$. The middle panel shows the cluster masses, in log scale. The bottom panel shows the distributions of the number of stars ``ejected'' from the clusters, $N_{ej}$, informed by \CMC. Both clusters experience episodic mass-loss, where the stars are preferentially stripped when the clusters are at their orbital pericenters, and a steady loss of mass along the rest of the orbit \label{fig:GC12_prop}}
\end{figure}

\subsubsection{Choosing example clusters}
For this work we chose two clusters (hereafter GC1 and GC2) that were accreted from an infalling dwarf galaxy and, at $z=0$, orbiting at the galactocentric distances similar to those of MW stellar streams believed to have GC progenitors. For the MW, these GC streams with surviving progenitors are located between 5.1 kpc \citep[NGC 6362;][]{sollima2020} and 31.6 kpc \citep[Pal 15;][]{myeong2017}. Both progenitors still exist at the present day with a significant fraction of their stars being stripped, and both are on eccentric orbits. 

Their basic properties as a function of time are shown in Figure \ref{fig:GC12_prop}. The top panel shows the 3D galactocentric physical distances, $r_{gal}$, of both clusters for the past $\sim2$ Gyr. GC1 is shown in blue, while GC2 is shown in red. Both clusters are on eccentric orbits, with the pericenter ($r_{peri}$) and apocenter ($r_{apo}$) distances slightly changing over time. We define the orbital eccentricity of the progenitor as, 
\begin{equation}\label{eq:e}
    \Tilde{e} \equiv \frac{r_{apo} - r_{peri}}{r_{apo}+r_{peri}}.
\end{equation} 
GC1 is on a less eccentric orbit with $(r_{peri},r_{apo})\simeq(6,15)$ kpc and $\Tilde{e}\simeq0.43$. GC2 is on a more eccentric orbit with $(r_{peri},r_{apo})\simeq(4,24)$ kpc and $\Tilde{e}\simeq0.72$. These values are consistent with MW GC streams (e.g., see Figure 7 of \cite{bonaca2025}). We note that although coming closer to the center of the host, GC2 is on an orbit with a longer orbital timescale compared to GC1.

The middle panel shows the clusters' masses as a function of time. Both clusters survive to $z=0$, with GC1 being slightly more massive than GC2. Both clusters continuously lose mass over the last two Gyr. However, their mass-loss rates are not continuous, implying that their mass-loss processes are not dominated by constant cluster evaporation.

The bottom panel shows the distributions of the number of escaped stars from each of the cluster, $N_{ej}$, as a function of time as determined by the collisional \CMC code. The escape distributions of both of the clusters are significantly non-uniform, with peaks coincide with steep declines in cluster mass (middle panel) and the orbital pericenters (top panel). The correlation with the orbital phases implies that both GC1 and GC2 primarily lose mass through tidal stripping. The stars inside the clusters are more likely to be stripped when the clusters are near their pericenters where their tidal radii are smallest. Table \ref{tab:GC12_props} summarizes the properties of GC1 and GC2.

\begin{deluxetable*}{lccccl}
\tablehead{
\colhead{GC} & \colhead{ $M_{init}$ ($\Msun$)} &\colhead{ $M_0$ ($\Msun$)} & \colhead{Age (Gyr)}& \colhead{$r_0$ (kpc)} & \colhead{$\Tilde{e}$}}
\startdata
GC1   & $4.45\times10^4$ & $3.15\times10^4$   &  10.12   & 9.4   & $0.43$   \\
GC2   & $2.14\times10^4$ & $1.48\times10^4$  & 7.94   & 22.5   & $0.72$  \\
\enddata
\caption{Properties of the example clusters: initial mass $M_{init}$, present-day mass $M_0$, age, present-day distance $r_0$, and orbital eccentricity $\Tilde{e}$ (Equation \ref{eq:e}). \label{tab:GC12_props}}
\end{deluxetable*}

\subsection{Step 3: Tracing the orbits of the escapers}
\begin{figure}
\plotone{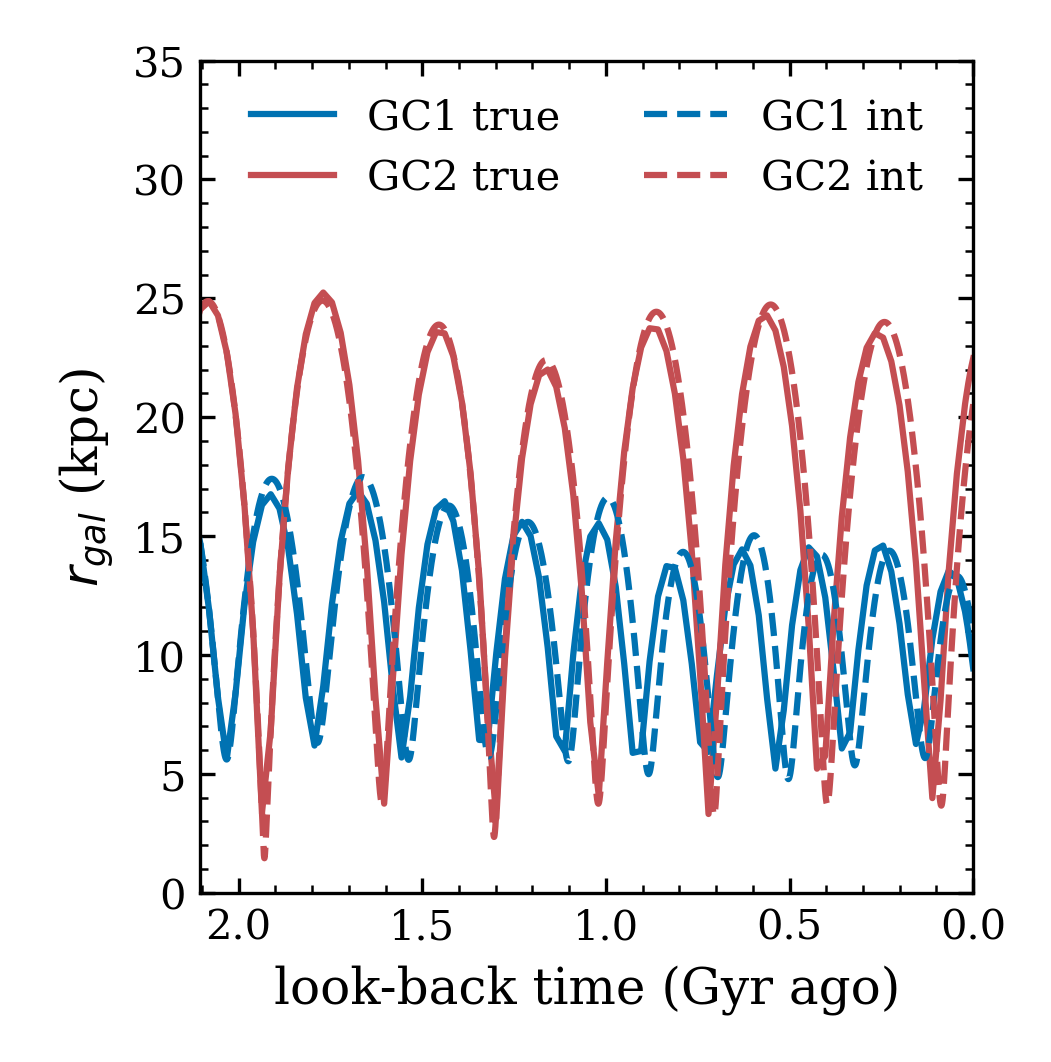}
\caption{Comparisons between the true (solid) and the integrated (dashed) orbits for GC1 and GC2 in blue and red, respectively. The solid lines are identical to the cluster orbits shown in Figure \ref{fig:GC12_prop}, while the dashed lines are the clusters' integrated orbits in the BFE potential model of our host galaxy\label{fig:GC12_orbits_true_vs_int}}
\end{figure}

We integrate the stripped stars generated by the globular cluster evolution model (Section \ref{subsec:gcformation}) in the combined host and cluster potential.

\subsubsection{Modeling the host galaxy}
We model the host galaxy with a time-dependent, non-parametric model of the simulated galaxy where the GCs were identified. This model captures the halo deformations and presence of massive Sagittarius-mass satellites \citep{arora2022, arora2025}. It is generated following the pipeline described in \citet{arora2022}. In brief, for each snapshot of the cosmological simulation we:
\begin{enumerate}
    \item Select all simulation particles (stars, gas, and dark matter) within 600 kpc of the center of the main halo;
    \item Model the combined potential of the dark matter and ``hot'' ($T>10^4$ K) gas using a spherical-harmonic expansion up to 8th order:
    \begin{equation}
        \Phi(r,\theta,\phi) \approx \sum_{\ell,m}^{\ell_{max}=8}\Phi_{\ell,m}(r)Y_{\ell}^m(\theta,\phi),
    \end{equation}    
    where $(r,\theta,\phi)$ are the spherical coordinates, $Y_{\ell}^m$ are the spherical harmonic function of the pole order $\ell$, $-\ell\leq m\leq\ell$, and $\Phi_{\ell,m}$ are the coefficients of the expansion given in logarithmically spaced bins in radius out to 300 kpc (approximately the present-day virial radius of the main halo);
    %The normalization of the density model is characterized using a spline in radius. 
    \item Model the combined potential of stars and ``cold'' ($T<10^4$ K) gas using a basis-function expansion to 8th order in azimuthal harmonics, with the potential in vertical and radial directions characterized using independent spline models with logarithmically spaced knots:
    \begin{equation}    \Phi(R,Z,\phi)\approx\sum_{m=0}^{m=8}\Phi_m(R,Z)\exp(im\phi),
    \end{equation}
    where $(R,Z,\phi)$ are the cylindrical coordinates with the galactic disk aligns with the $Z=0$ plane.
\end{enumerate}
These models are constructed using \textsf{agama} \citep{vasiliev2019} in the frame centered on the host galaxy center (the ``galactocentric frame'') in each snapshot, which is not an inertial frame. We model the acceleration of the host galaxy's center of mass over time using independent splines for each Cartesian coordinate in the simulation box. We fix the orientation of the galactocentric frame so that the $Z$ axis is perpendicular to the disk plane at the present day. In practice, the disk plane is not entirely stable but for this simulated galaxy it changes by less than 20 degrees over the integration time for the stripped stars. This is well within the grid used for the spline model of the disk. This strategy reproduces instantaneous gravitational forces to within a few percent over the entire simulated galaxy, and reproduces halo-like orbits over multiple dynamical times \citep{arora2024}.

For validation, we forward integrated the orbits of the progenitors of GC1 and GC2 for 2.1 Gyr up to $z=0$ using our constructed host galaxy model. We interpolate the BFE forces for the integration linearly between snapshots, as discussed in \citet{arora2024}. This integration time corresponds to 100 simulation snapshots. The 6D initial conditions of both clusters and their ``true'' orbits were taken directly from their corresponding tracer particles from the FIRE snapshot data. Figure \ref{fig:GC12_orbits_true_vs_int} shows the comparisons between the true orbits from the snapshot data (identical to the top panel of Figure \ref{fig:GC12_prop}) and the integrated orbits in the BFE potential model. We can reconstruct the orbits of GC1 and GC2 reasonably well. Although GC1 has a larger $r_{peri}$ and is less eccentric, it has a shorter dynamical timescales between each orbit, relative to the time between simulation snapshots, which results in worse orbit reconstruction. We see slight orbital phase shifts for both of the clusters and the shift is increasing with time for GC2. However, we note that our model produces consistent values of $r_{peri}$ and $r_{apo}$ for both clusters. Even for GC1, the $r_{apo}$ mismatch is $\lesssim1$ kpc.

\subsubsection{Modeling the cluster potential}
The orbital evolution of each of the escaped star not only depends on the potential of the host galaxy, but also on the cluster's own self-gravity. This is because \CMC immediately removes the stars once they meet the stripping criteria (i.e., $r_{apo}^{GC}>r_{tidal}$), not when they actually cross the cluster's tidal boundary. The cluster model is spherically symmetric, so in order to ensure that the stars leave from the Lagrange points of the cluster we model it as a Plummer sphere with evolving mass and scale radius informed by the \CMC model of the progenitor. Specifically, the cluster potential is given by,
\begin{equation}
    \Phi_{GC}(r,t) = -\frac{GM_{GC}(t)}{\sqrt{r^2+a_{GC}^2(t)}},
\end{equation}
where $M_{GC}(t)$ and the $a_{GC}(t)$ are the mass and the scale radius of the cluster at simulation time $t$, respectively.
%\NCW{[See comment in code editor (commented out due to length).]}
%\NCW{[As a reminder, a single Plummer potential fit will significantly underestimate the depth of CMC's potential well in the cluster core, causing spuriously high ejecta energies when evolving their trajectories with a Galactic dynamics integrator \citep[e.g., Figure 2 of][]{weatherford2024}, and inflated stream width and velocity dispersion. This was a big issue for me when modeling CMC streams on circular GC orbits, since about half of the escapers in that case are removed from CMC within the core radius. I fervently hope in your case that the enhanced stripping at perigalacticon means that most stars are instead removed from CMC outside the core radius, ensuring that the underestimate of the potential depth in the core does not cause too dramatic a bias toward higher ejecta energy. To double-check, can you perhaps plot (maybe share in Slack channel) the distribution of ejecta's radial position in the cluster, normalized by the cluster's core radius, at the time of their removal from CMC? If a significant fraction are removed from CMC within the core radius, you then may want to double-check how well your Plummer potential is fitting the true potential in snapshots from your CMC simulations.]}
%\FIXME{fix this if we actually use the Zhao's bases}

\subsubsection{Stream formation, and orbital integration}
\label{subsec:streamform}

We choose all stars that have met the stripping criterion from the GC in the past 2 Gyr according to the \CMC model because --- 1) we find that stars ejected further in the past than this tend to be relatively hot and/or phase-mixed, and 2) the difference in orbital phases between the true and integrated orbits grow over time as shown in Figure \ref{fig:GC12_orbits_true_vs_int}. For our two example streams this interval corresponds to about the last $\sim~10$ pericenter passages of the GC.  The stars are injected into the BFE model at the distance and speed relative to the cluster provided by \CMC. The ejection radius and speed are located inside the cluster, so we choose the angular position and velocity direction from isotropic distributions and allow the stars to escape the cluster potential model. 

Once the stars have escaped, we continue integrating them in the BFE for the remaining time, taking into account different stripping times, until the end of the simulation (the ``present day'').

\begin{figure*}
\plotone{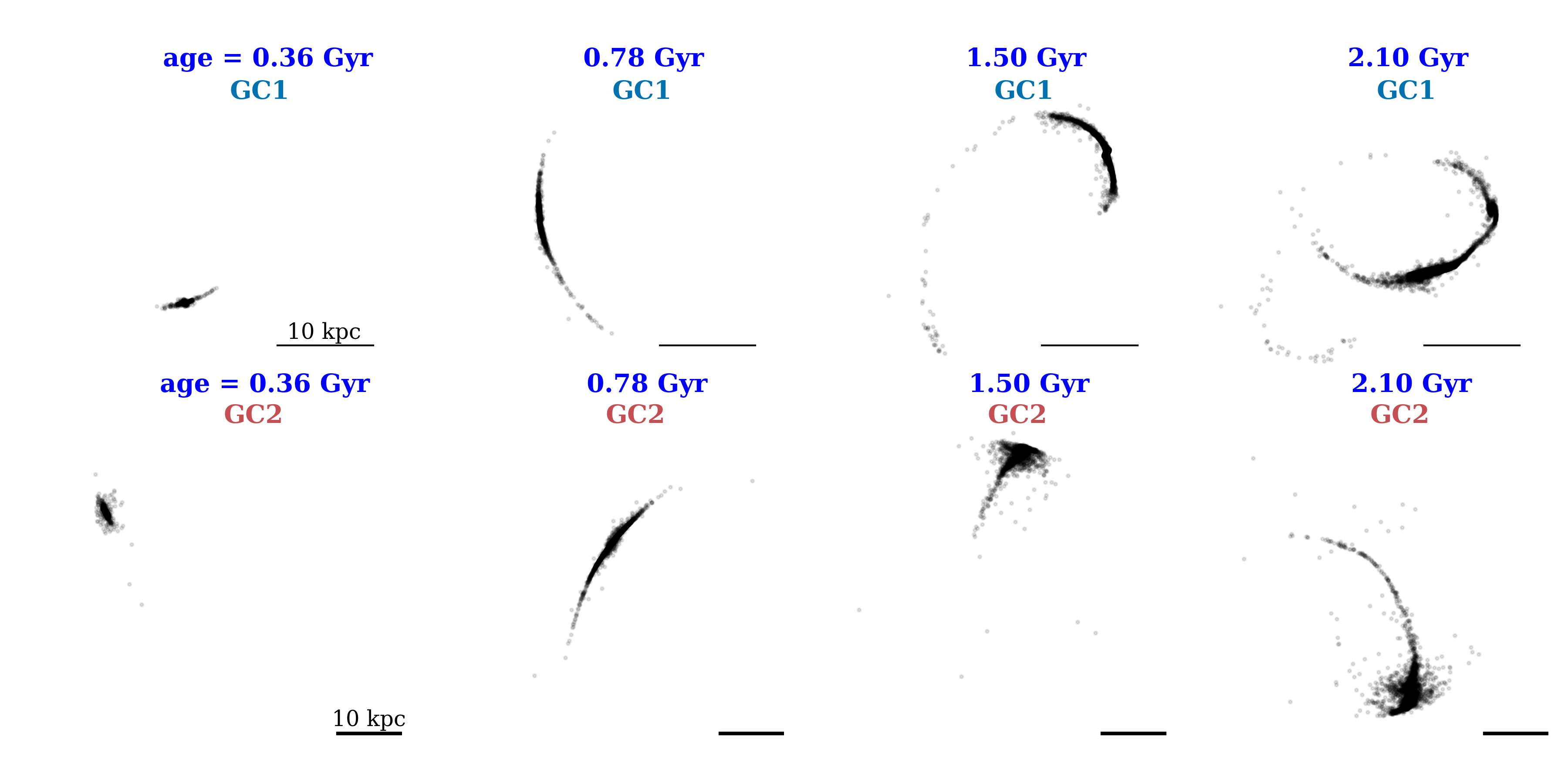}
\caption{GC1 and GC2 streams (44045 and 19206 stars, respectively) shown in 4 different age bins: 0.36, 0.78, 1.50, and 2.10 Gyr. GC1, on a less eccentric orbit, forms a relatively cold and long stream. GC2, on a more eccentric orbit, forms a shell-like structure. \label{fig:GC12_streams}}
\end{figure*}

\subsection{Terminologies: $t_{ej}$ vs. $t_{escp}$}\label{subsec:t_ej_vs_t_escp}
In GBoF2, \CMC uses a simple radial criterion, where in the cluster-centric frame a star with $r_{apo}^{GC} > r_{tidal}$ is considered ``stripped''. We use $t_{ej}$ to refer to the time when a given star meets the radial stripping criterion. This also corresponds to the time when we inject the star into our orbit integration pipeline (i.e., the ``initial conditions'') described in Section \ref{subsec:streamform}. Most of the stars are still well within the cluster at $t_{ej}$.

As a result, there will be a delayed time between the $t_{ej}$ and when the star actually crosses the tidal boundary and escapes the cluster. It has been shown for most stars where their energies is just slightly above the escape energy from the cluster, they cannot escape the cluster easily unless they pass through one of the two Lagrange points where the energy barrier is lowest \citep[e.g.,][]{fukushige2000, baumgardt2001,weatherford2024}. Moreover, for an eccentric cluster orbit, $r_{tidal}$ fluctuates along the orbit. Therefore, some of the stars that have already escaped may still be recaptured due to this time-dependent nature of $r_{tidal}$. 

Since our goal is to study the overall morphology and the properties of the tidal tails, all of which are at scales of a few kpc, we use a simple distance criterion to define the ``escape time''. Specifically, we say that the escape time $t_{escp}$ corresponds to the time when the star first crosses a pre-determined distance threshold away from the cluster $r_{\text{threshold}}$. Since GC1 and GC2 have roughly similar masses, we set $r_{\text{threshold}}=100$ pc, i.e.,
\begin{equation}
    r\rvert_{t=t_{escp}} = 100\ \text{pc}.
\end{equation}
The cluster's tidal radius can be approximated by the instantaneous eigenvalue of the tidal tensor \citep[e.g., see Equation 10 in][]{renaud2011}:
\begin{equation}\label{eq:r_tidal}
    r_{tidal} = \left(\frac{GM_{GC}}{\lambda_{e,1}} \right)^{1/3},
\end{equation}
where $M_{GC}$ is mass of the cluster, and $\lambda_{e,1}$ is the largest eigenvalue of the tidal tensor. We note that $r_{tidal}$ is both a function of time and location of the cluster within the host. For both of our clusters, $r_{tidal}$ is strictly less than 100 pc, ensuring that the stars that have met the condition shown in Equation \ref{eq:r_tidal} have truly escaped and not recaptured by the fluctuations in the cluster's $r_{tidal}$.

%For a system with $(R_{GC}, m_{GC}, M_{gal})=(15\ \text{kpc}, 10^{4}\Msun, 10^{12}\Msun)$, the tidal radius is $\lesssim10$ pc, much less than the distance threshold of 100 pc to ensure that these escaped stars are not recaptured.

To summarize, $t_{ej}$ refers to the time when each star meets the radial stripping criterion set by the \CMC and this sets the initial conditions where we start orbit integrating the star in the combined host and cluster potential. This is not to be confused with $t_{escp}$ which refers to the time when each star crosses the distance threshold of 100 pc away from the cluster progenitor.

\section{Results} \label{sec:results}
The GC1 (top) and GC2 (bottom) streams are shown in Figure \ref{fig:GC12_streams} at four different times, when they are 0.36, 0.78, 1.50, and 2.10 Gyr old. The right column corresponds to the end of the simulation (the present day). These are 2D projected scatter plots in a rotated galactocentric frame where each stream appears face on. In total, we perform orbit integration for 44045 and 19206 stars in GC1 and GC2, respectively.

At the present day, GC1, on a less eccentric orbit, forms a stream that is relatively long ($\sim120^\circ$) and thin, with significant variations in density along it. This stream is also cold, with velocity dispersions of less than 5 km \unit{s}{-1}, consistent with recent measurements of streams in the MW that are believed to originate from GC progenitors \citep[e.g.,][]{ibata2017, price-whelan2019, gialluca2021, li2022, kuzma2022, valluri2025, tavangar2025}. Interestingly, the velocity dispersion is lowest near the progenitor and increases slightly ($\sim$ 5 km/s) toward the ends of the stream, suggesting that heating of the stream stars from one of several processes is dominating over gravitational cooling due to conservation of phase space volume (i.e. Liouville's theorem). We explore this further in Section \ref{subsec:dispersion}. 

At the present day, GC2, on a more eccentric orbit, forms a shorter stream that contains both a thin stream-like segment and a diffuse shell-like segment. However, we note that the vast majority of the GC2 stream stars belong to the diffused component. The thin streams only consists of a few hundred stars ($\sim1-2\%$) that were ejected from the cluster over 1 Gyr ago. The orbital plane of the GC2 progenitor orbit is also visibly changing.

\subsection{Velocity dispersion}\label{subsec:dispersion}

\begin{figure*}
\plotone{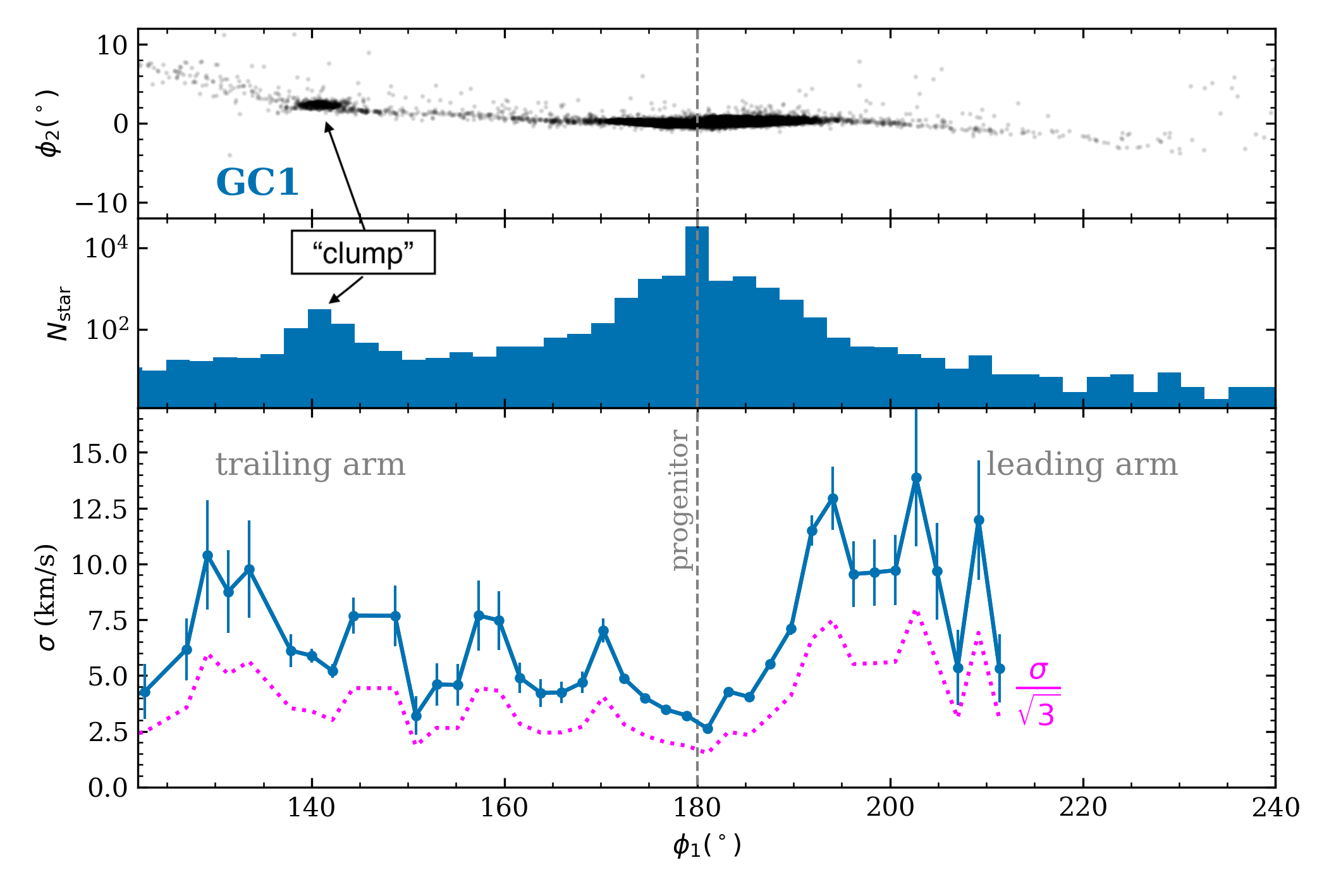}
\caption{Top: The GC1 stream at the present day in stream-aligned coordinates. There is an asymmetry in the lengths of the leading and trailing arms.
Middle: The $\phi_1$ distribution of all stream stars.
Bottom: The 3D velocity dispersion, $\sigma$, along the stream, binned uniformly in $\phi_1$ with a bin size of approximately 2$^\circ$. The dotted line (magenta) shows the estimated 1D velocity dispersion. $\sigma$ is lowest near the progenitor’s location and increases as one moves away from it. The dispersions in $\sigma$ were estimated assuming Gaussian velocity distributions within each bin. \label{fig:dispersion}}
\end{figure*}

It is interesting to speculate the stellar velocity dispersion along a given stream since this quantity is often used as a proxy for the stream width and the progenitor mass. 

To estimate the velocity dispersion along GC1, we first transform the GC1 stream into its stream-aligned coordinate system $(\phi_1, \phi_2)$. Specifically, we assume the observer is located at the center of the host galaxy and compute the total angular momentum vector of the GC1 progenitor, $\vec{L}_{GC1}$, at the present day. We then rotate to a new coordinate system in which $\vec{L}_{GC1}$ points along the Z-axis and shift the progenitor to be at $\phi_1 = 180^\circ$. The top panel of Figure \ref{fig:dispersion} shows the GC1 stream in this coordinate frame. The leading arm has $\phi_1 > 180^\circ$ and the trailing arm has $\phi_1 < 180^\circ$. We also see asymmetry in the length of the streams where the trailing arm is much more extended than the leading arm.

We bin the GC1 stream along $\phi_1$ into equal segments of $2^\circ$ each. Within each two-degree stream segment, we select a subset of stars within the one-sigma scatter around the local median $\phi_2$ and compute the 3D velocity dispersion, $\sigma$, among these stars for all the stream segments that contain at least 10 stars. This selection in $\phi_2$ is to remove outlier stars that locate far from the main stream track. 

The bottom panel of Figure \ref{fig:dispersion} shows the 3D velocity dispersion, $\sigma$, along $\phi_1$ for the GC1 stream. The dispersions in $\sigma$ were estimated assuming Gaussian velocity distributions within each bin. The dashed vertical line marks the location of the GC1 progenitor. $\sigma$ is lowest near the progenitor with $\text{min}(\sigma)\simeq 2.6$ km \unit{s}{-1} and it is increasing along both the leading and trailing arms. This increase implies that dynamical heating from other processes, such as overlapping of stream stars escaped at different times and interactions with other substructures, is dominating over the gravitational cooling due to conservation of phase space volume implied by Liouville's theorem. Stream stars that escaped from different stripping episodes can overlap and inflate the local dispersion. This is especially true for regions near the progenitor where multiple stripping episodes overlap. On the other hand, we see evidence of the decrease in $\sigma$ as we go sufficiently away from the progenitor ($\phi_1<135^\circ$ or $\phi_1>200^\circ$) where there are none or few overlaps. Moreover, although being truncated at $\ell\leq8$, our BFE galactic potential model contains substructures such as subhalos. Massive subahalos with masses similar to Sagittarius dwarf galaxy are present even in the $\ell=2$ mode \citep[see Figure 7 in][]{arora2022}.
%set up a grid in galactic latitude ($b$) and galactic longitude ($l$) with equal spacings in $\cos b $ and $l$. We denote the new pole location ($b_{pole}$,$l_{pole}$) that minimizes 

For GC2, it is more difficult to transform into its stream-aligned coordinate system. Due to its highly eccentric orbit, most of the stream stars in the GC2 stream reside in the diffuse shell-like component and only $\sim1-2\%$ of the escaped stars make up the thin stream-like component. For the stream-like component, the typical value of the local dispersion among five nearest neighbors is $\sigma_{\text{stream}}\approx10$ km \unit{s}{-1}. For the shell-like component, the total dispersion computed among all the stars is $\sigma_{\text{shell}}\approx26$ km \unit{s}{-1}.

\subsection{Stream substructure}

The middle panel of Figure \ref{fig:dispersion} shows the $\phi_1$ distribution of all the GC1 stream stars at the present day with the progenitor location marked by the dashed vertical line. We see that the stellar number density is highest at the location of the progenitor, $\phi_1 = 180^\circ$. This is so for two reasons. First, some of the stars were recaptured due to the fluctuations of the tidal boundary of the cluster (see Section \ref{subsec:t_ej_vs_t_escp}). $r_{\text{tidal}}$ shrinks and expands over time as all of these quantities are time-dependent, especially for an eccentric orbit. Second, the regions closest to the progenitor are dominated by the group of stars that were ejected in the most recent stripping episode. For GC1, almost $\sim50\%$ of the stream stars were ejected during this episode (see the bottom panel of Figure \ref{fig:GC12_prop}).

The stellar number density generally decreases as we move further away from the progenitor. However, the GC1 stream contains a clear local stellar overdensity (hereafter, the ``clump'') in the trailing arm around $\phi_1=140^\circ$. We briefly discuss possible origins of the clump below.

\begin{figure}
\plotone{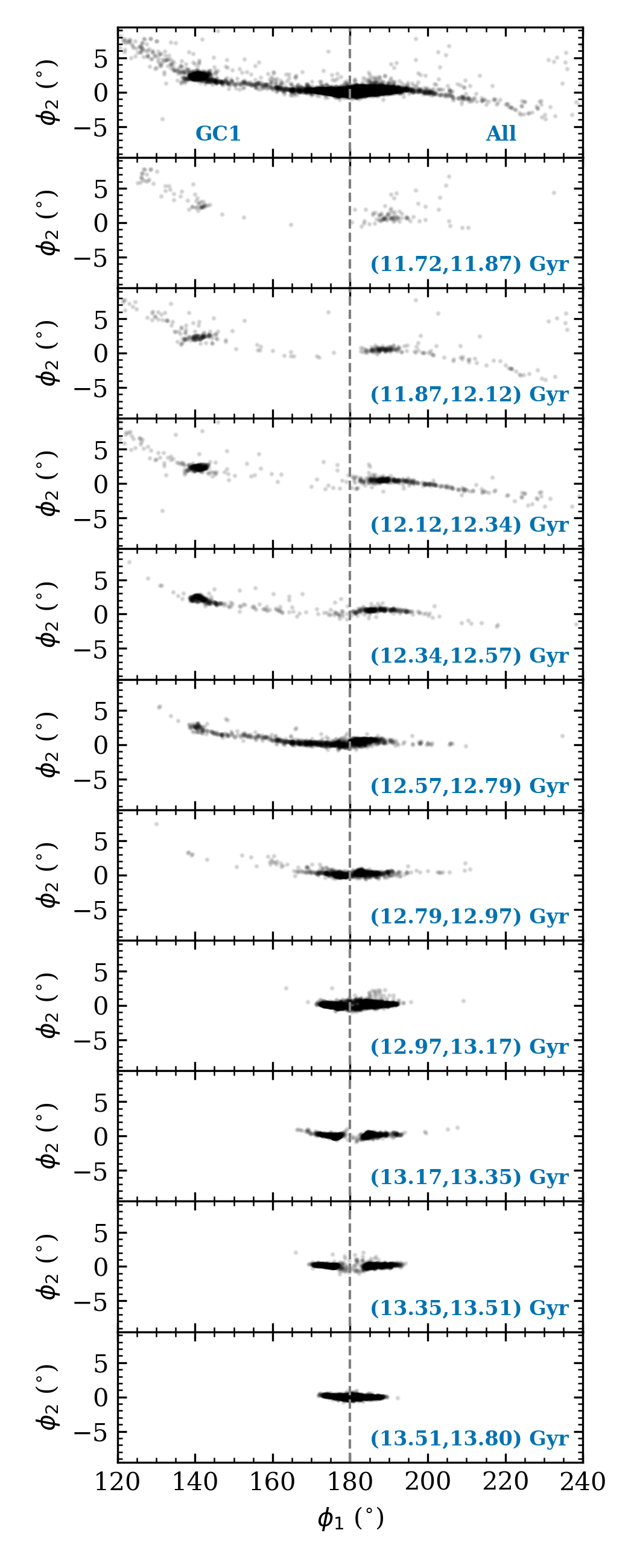}
\caption{Subsets of the GC1 stream stars at the present day in the ($\phi_1$, $\phi_2$) coordinates grouped by their escape times, $t_{escp}$. All of the stream stars are plotted in the top panel (identical to Figure \ref{fig:dispersion}). In the remaining panels, only subsets of stars with escape time $t_{escp}$ between the times specified by the blue numbers in the bottom right of each panel are shown. In total, GC1 went through 10 different stripping episodes over $\sim2$ Gyr. \label{fig:GC1_stripping_episodes}}
\end{figure}

Figure \ref{fig:GC1_stripping_episodes} shows the present-day GC1 stream, with subsets of stars grouped by their stripping episodes. The top panel displays all the stream stars, while the remaining panels show only those stars escaped within the time ranges indicated by the blue numbers in the bottom right of each panel (see the definition of $t_{esp}$ in Section \ref{subsec:t_ej_vs_t_escp}). These time ranges were selected so that each panel contains stars stripped during the same pericentric passage, ordered from the earliest to the most recent. Interestingly, the clump includes stars from multiple stripping episodes, all of which were ejected between $\sim1-2$ Gyr ago. Since these stars escaped the cluster at significantly different times and with varying initial properties, this suggests that the clump is unlikely to have originated from the initial dynamical conditions of the stream stars at the time of their escape, but rather from physical processes specific to this particular stream orbit such as interactions with other Galactic substructures \citep[e.g.,][]{amorisco2016,erkal2017,pearson2017,bonaca2020gd1sgr,nibauer2025}. 

One possible explanation is an interaction with galactic substructure, such as a subhalo or the Galactic disk—particularly given that the GC1 orbit is highly aligned with the disk plane (albeit in retrograde motion relative to the disk stars). Theoretically, streams in retrograde orbits experience fewer perturbations \citep{hattori2016,pearson2017}. However, snapshots from the FIRE simulation show evidence that the disk was perturbed by a massive $10^{10}\Msun$ subhalo that reached a pericenter distance of $\sim45$ kpc around 1 Gyr ago. Interestingly, this timing coincides with the escape times of the stars in the clump (i.e., it contains only stars that escaped the cluster over 1 Gyr ago). Moreover, the stars in the overdensity clump are near the apocenters of their orbits, which further enhances the clump’s visibility due to a pile-up at apocenter. Further investigation is required to confirm the origin of the clump and the general formation of such feature. This will be explored in future work.

GC2, which is on a more eccentric orbit, produces a unique stream morphology that includes both a thin, stream-like structure and a diffuse, shell-like structure. Interestingly, observations of Milky Way stellar streams also reveal that some exhibit similar two-component morphologies, with both thin and diffuse features \citep[e.g., Jhelum,][]{bonaca2019}. The shell-like structure is expected for highly eccentric stream orbits, as they primarily phase-mix due to the spread in orbital angular momentum \citep[e.g.,][]{hendel2015}; however, this alone cannot explain the presence of the thin component.

\begin{figure}
\plotone{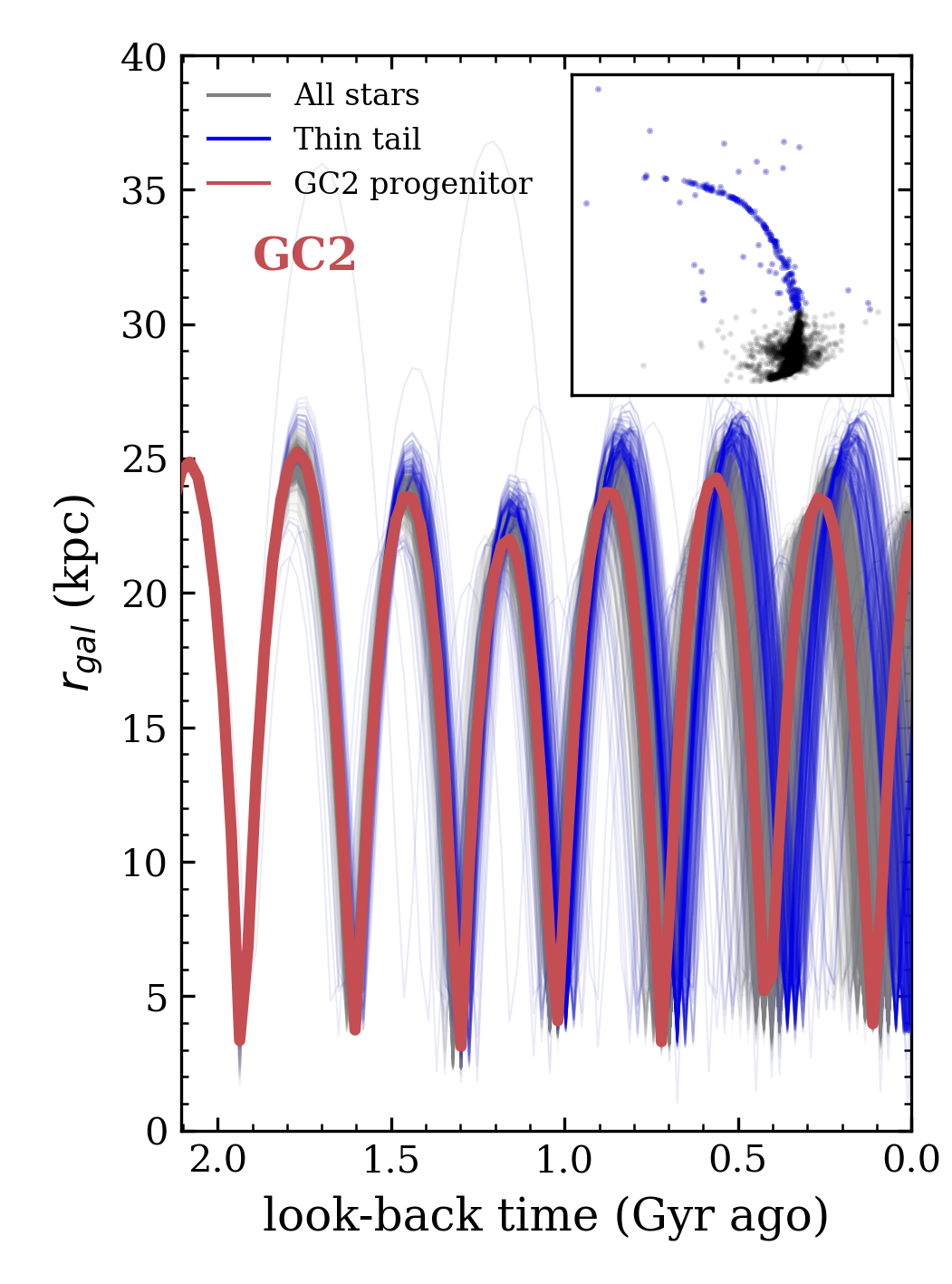}
\caption{Orbits of the GC2 stream stars and its progenitor. The orbits of all the stream stars, stars that comprised the thin tail, and the progenitor are shown in gray, blue, and red, respectively. At the present day, the progenitor is located near the apocenter, while the stars belonging to the thin tail. \label{fig:GC2_orbital_phase}}
\end{figure}

For GC2, the thin and diffuse components naturally arise from differences in the orbital phases of the GC2 stream stars. Figure \ref{fig:GC2_orbital_phase} shows the orbits of the GC2 stream stars and the GC2 progenitor. The orbits of all stream stars, those that comprise the thin tail at the present day, and the progenitor are shown in gray, blue, and red, respectively. The inset scatter plot in the upper right illustrates the selection of the thin tail. The stars in the thin tail have slightly longer orbital timescales compared to the progenitor, which contributes to accumulated phase differences over time. At the present day, the progenitor is located near its slow-moving apocenter, resulting in a visible pile-up of stars. In contrast, the stars in the thin tail are near their fast-moving pericenter.

\subsection{Stream orbital pole}
A stream orbit in a time-independent, spherical Galactic potential model is confined within a plane. This is because the total angular momentum is conserved. For an axisymmetric potential, only the Z-direction of the angular momentum is conserved, so the stream orbit will wobble around the symmetry axis at fixed frequencies. However, this is not the case for cosmological orbits where the time-dependent and triaxial nature apply the time-dependent gravitational torque that causes the orbital angular momentum vector to precess chaotically.

\begin{figure}
\plotone{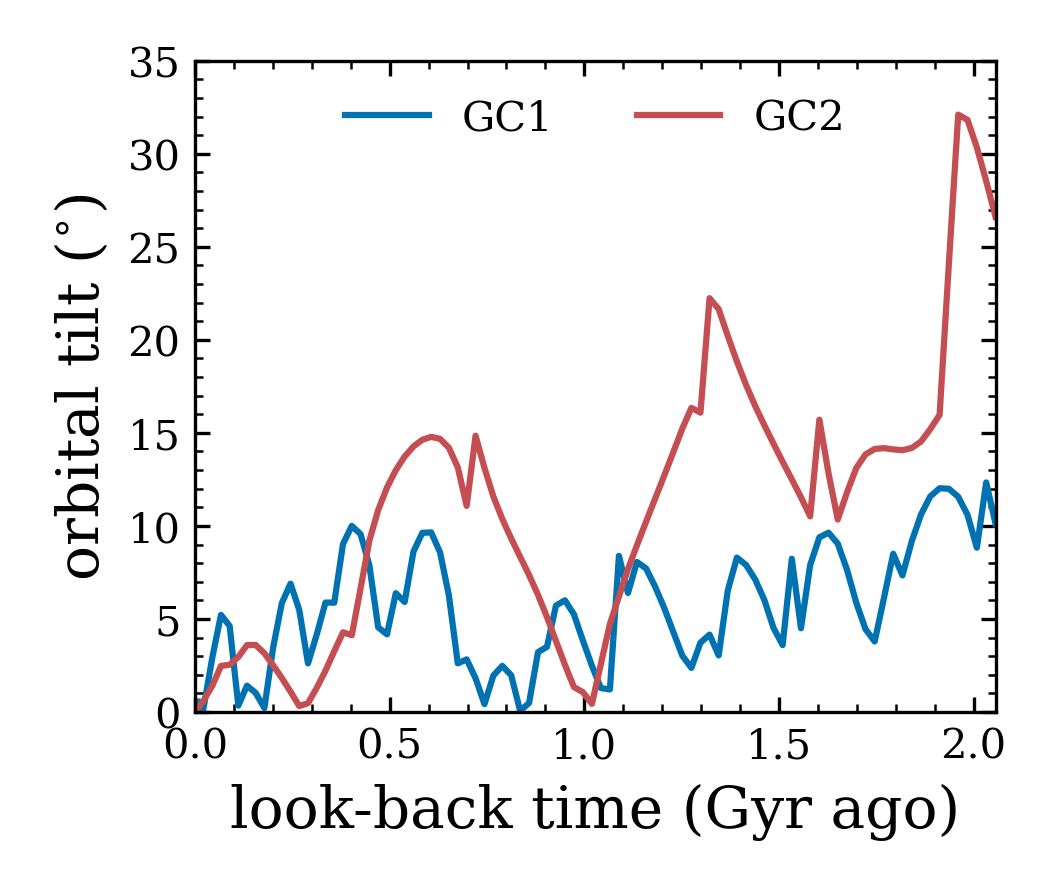}
\caption{The precession of GC1 and GC2's orbital planes over time. The orbital tilt is define as the angle between the total angular momentum vector of the progenitor at time $t$ and at the present day. \label{fig:GC12_orbital_tilt}}
\end{figure}

Both GC1 and GC2 show considerable non-periodic precession of their orbital planes over time. Figure \ref{fig:GC12_orbital_tilt} shows the tilt angle as a function of look-back time. The orbital tilt angle, $\theta_{\text{tilt}}$, is defined as the angle between the direction of the total orbital angular momentum vector of the progenitor at any given time, $\hat{L}_{\text{GC}}$, and at the present day, $\hat{L}_{\text{GC},0}$:
\begin{equation}
    \theta_{\text{tilt}} = \cos^{-1}(\hat{L}_{\text{GC}}\cdot\hat{L}_{\text{GC},0}).
\end{equation}

The orbital plane of the GC1 progenitor orbit changes by $>10^\circ$, while this change is more extreme at $>30^\circ$ for GC2. This precession of the stream stars' orbital planes causes the stream to widen \citep{erkal2016,dehnen2018} and, in some cases, produces an observable orbital-phase dependent divergence of the stream track (Section \ref{sec:div_stream_track})

\section{Discussion} \label{sec:discussion}
In Section \ref{sec:results}, we present two example cosmological GC streams from our model, both of which exhibit interesting morphology at the present day. In this section, we focus on the interesting time-dependent stream features as well as the observational implications.

\subsection{Orbital-phase dependent stream track misalignment}\label{sec:div_stream_track}
It has been shown that stellar streams, whether on circular or eccentric orbits, contain substructures that resemble``streaks'' or ``feather'' \citep[e.g.,][]{kupper2008, just2009, kupper2012, amorisco2015, fardal2015}. However, the origins of these feathers are fundamentally different. In the circular case, they arise from epicyclic oscillations; thus, the substructures along a given stream are driven by the current orbital phases of the stream stars \citep{kupper2008, just2009}. In contrast, for highly eccentric orbits, the cluster’s mass loss is episodic (as seen in GC1 and GC2), so the feathers instead originate from the orbital phases of the stream stars at the time of ejection \citep{amorisco2015, fardal2015}. Despite the predictions that these feathers exist, MW GC streams are often represented by only their average stream tracks \citep[e.g.,][]{mateu2023}. This may be due to a combination of on-sky projection effects and the lack of high-precision 6D phase-space data for stream stars. This motivates the need to fully characterize these feathers, examine their time evolution, and how they affect the average stream track.

The GC1 stream shows clear feathering around the cluster's location, each consisting of stars with similar $t_{escp}$. In stream-aligned coordinates, the GC1 stream exhibits a well-defined track at the present day (top panel of Figure \ref{fig:dispersion}). Since this is a simulated stream, it is possible to examine its track over time.

\begin{figure*}
\plotone{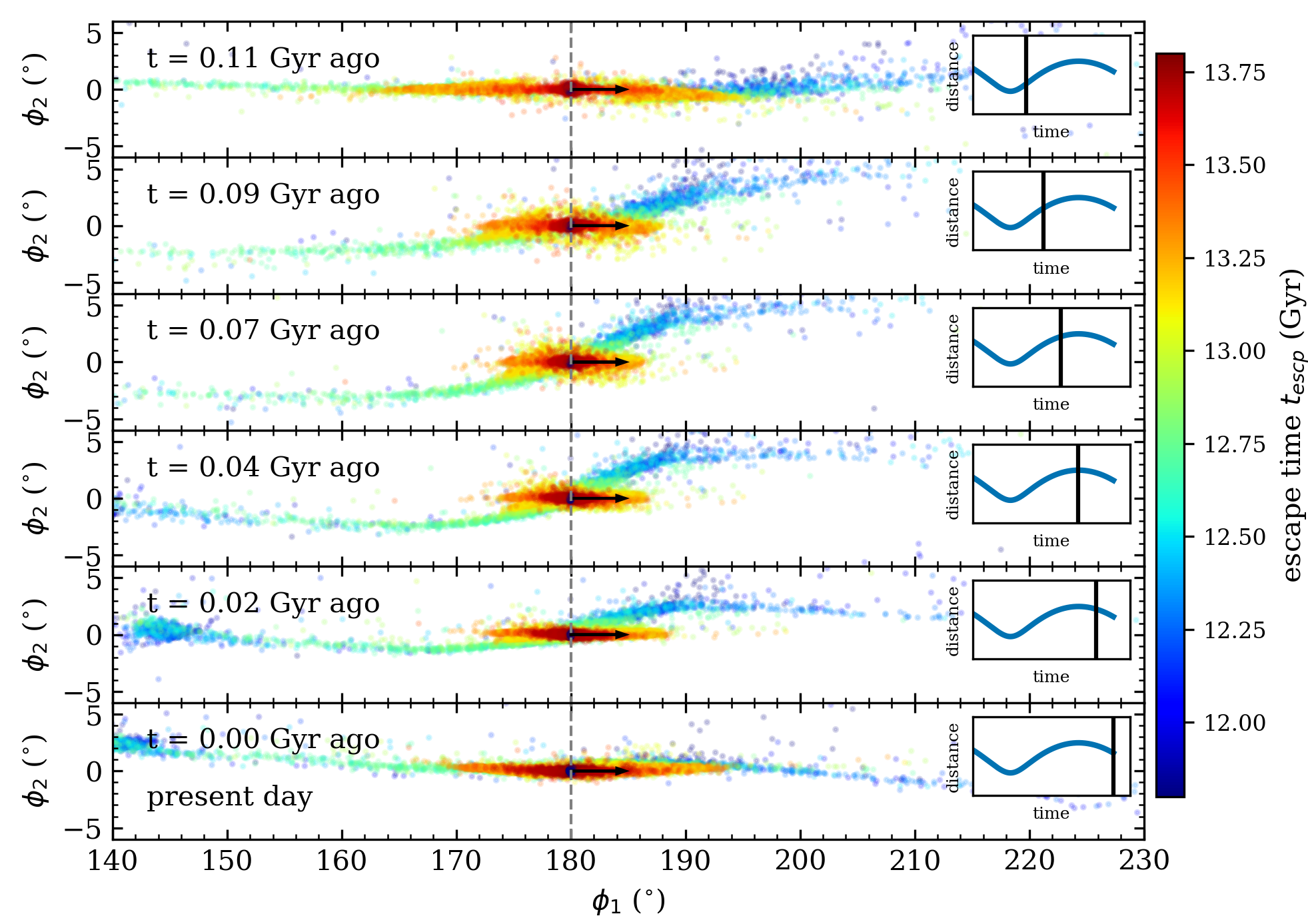}
\caption{The orbital-phase dependent misalignment of the stream track. Each panel shows the GC1 stream in the stream-aligned coordinates in different time bins. The progenitor is located at $\phi_1=0^\circ$ and its velocity vector is shown by the black arrow. The color represents the time that each star escaped from the cluster, $t_{escp}$. The inset plots on the right show the orbital phases of the progenitor. The stream track traced by the stars across different $t_{escp}$ is decently aligned in the top and the bottom panels where the progenitor is near pericenter (either having slightly moved past it or is approaching it). However, we see a significant misalignment of the stream track around the progenitor near apocenter due to the orbital mismatch between the most recent and the earliest stripped stars. \label{fig:div_stream_track}}
\end{figure*}

Figure \ref{fig:div_stream_track} shows the GC1 stream in the instantaneous stream-aligned coordinate frame across six time bins spanning the last 110 Myr. This period represents roughly half of the stream’s orbital period, capturing its state near both pericenter and apocenter. The progenitor is fixed at $\phi_1 = 180^\circ$ in all panels, with its velocity pointing rightward along the $\phi_2 = 0^\circ$ axis, as indicated by the arrow. Each stream star is colored according to its ejection time from the cluster, $t_{ej}$: red stars were ejected most recently, while blue stars were ejected long ago. The inset plot on the right of each panel indicates the progenitor's location along its orbit — i.e., its orbital phase — at each corresponding time. The bottom panel corresponds to the present day.

The stream track traced by stars near the progenitor is well aligned across all $t_{escp}$ in the top and bottom panels, where the progenitor is near pericenter --- either having just passed pericenter or approaching it, respectively. In contrast, the middle four panels show visible misalignments in the stream track near the progenitor to varying degrees. These middle panels correspond to when the progenitor is near apocenter. The misalignment is orbital-phase dependent and is most pronounced when the progenitor is exactly at apocenter, at $t = 13.75$ Gyr. 

For the GC1 stream on an eccentric orbit, this behavior can be explained by orbital-phase compression and stretching near apocenter and pericenter, respectively. Near pericenter, stream stars experience increased and rapidly changing orbital (angular) velocities, which stretch the stream and reduce the degree of stream track misalignment.

The stream track misalignment is driven by the orbital mismatch between the most recent and the earliest stripped stars as pericenter distance, apocenter distance, and the orbital plane of a cosmological orbit can change significantly over time (Figure \ref{fig:GC12_prop} and Figure \ref{fig:GC12_orbital_tilt}). As expected, the present-day progenitor's velocity vector is well-traced by recently escaped stars (red), but not so much by stars that have escaped the cluster over $\sim1$ Gyr ago (green and blue). Finally, we note that this stream track misalignment becomes much more obvious for GC1 in its last orbit since nearly half of its stream stars were ejected during the last pericentric passage alone (Figure \ref{fig:GC12_prop}). Therefore, the misalignment effect could be much less obvious if the cluster mass-loss is highly dominated by a single stripping event. 

Generally speaking, stream stars do not delineate a single orbit, and a misalignment between the direction in which the stream grows (i.e., the stream track) and the progenitor’s velocity vector is expected \citep{eyre2011, sanders2013a, sanders2013b}. This behavior is naturally explained within the action-angle-frequency framework. In this context, the Hessian matrix --- i.e., the matrix of second partial derivatives of the Hamiltonian --- 
is typically dominated by a single eigenvalue, and the stream will grow along the direction of the corresponding eigenvector. This direction is, in general, misaligned with the progenitor’s frequency vector. However, this misalignment is usually small and, more importantly, independent of escape time for a stream formed in a time-independent, axisymmetric Galactic potential.

\subsection{Mass segregation}
\cosmogems produces GC streams composed of individual stars, each with their own masses, rather than identical-mass star particles (refer to Section \ref{subsec:gcformation}). This allows us to investigate potential mass-dependent trends in stellar ejection in our two example clusters, and how these trends may influence the present-day structure of GC streams.

One well-known dynamical process within GCs is mass segregation --- a phenomenon in which more massive objects, such as high-mass stars and black holes, tend to sink toward the cluster center, while lower-mass stars migrate outward \citep[e.g.,][]{spitzer1987book,bonnell1998}. The black holes are usually much more massive than the stars and will sink first. Upon the ejection of the central black hole population through strong encounters, the GC becomes core-collapsed, transitioning from a flat to a steep central density profile \citep[e.g.,][]{kremer2018,kremer2019}. \citet{balbinot2018} used the synthetic cluster evolution code \textsf{EMACSS} \citep{gieles2014, alexander2014}, coupled with a semi-analytic model for the evolution of the stellar mass function, and found that stellar streams are more likely to be detected near the time of cluster dissolution due to a higher proportion of massive stars escaping.

\begin{figure*}
\plotone{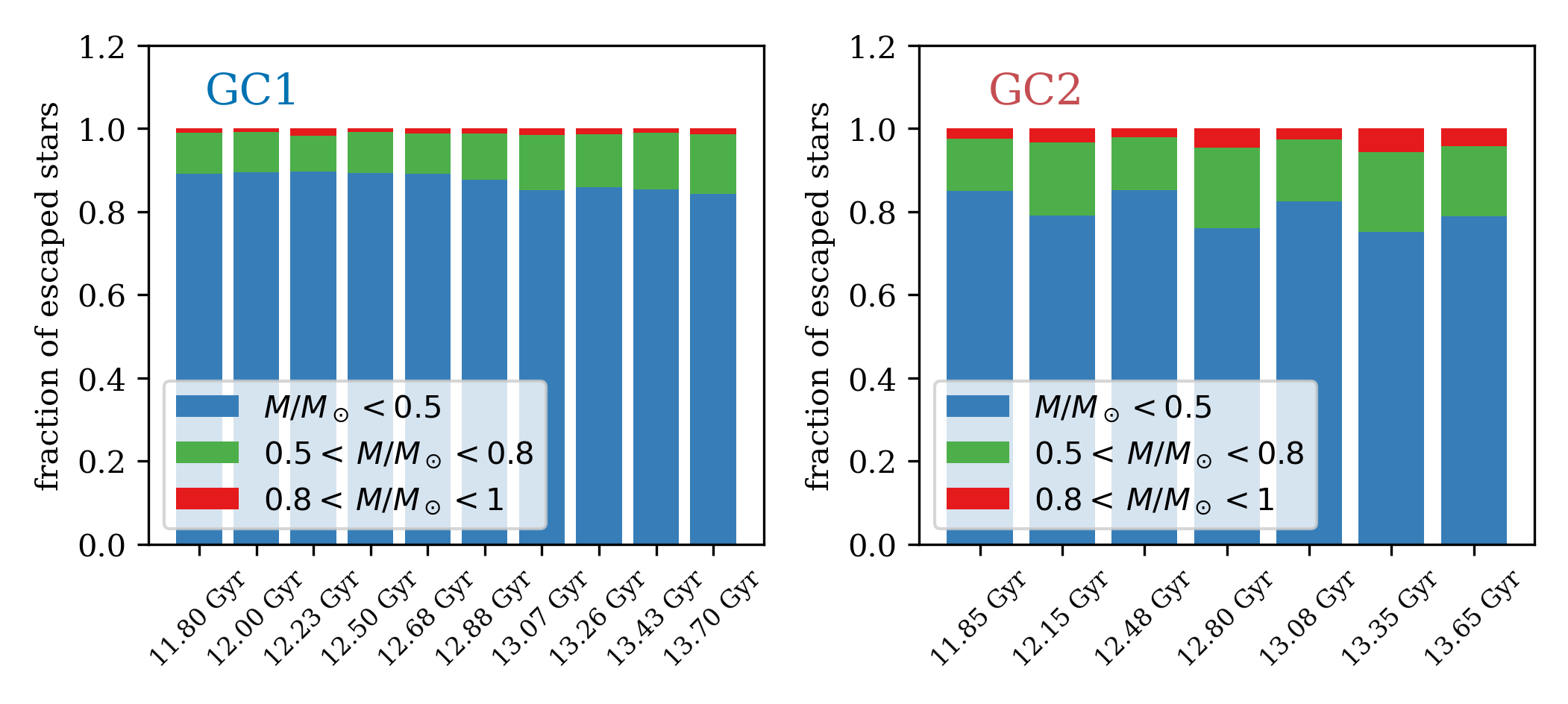}
\caption{Bar charts showing the fraction of escaped high-mass (red; $0.8 < M/\Msun < 1.0$), intermediate-mass (green; $0.5 < M/\Msun < 0.8$), and low-mass (blue; $M < 0.5\Msun$) stars in each episode. The left panel displays the data for GC1 (10 stripping episodes), while the right panel shows the data for GC2 (7 stripping episodes). The x-axis labels indicate the average time of each stripping episode. \label{fig:GC12_stripping_bar_chart}}
\end{figure*}

For GC1 and GC2, we do not observe significant variation in the fractions of high-mass and low-mass escapers across different stripping episodes. Figure \ref{fig:GC12_stripping_bar_chart} presents bar charts showing the fraction of escaped high-mass (red; $0.8 < M/\Msun < 1.0$), intermediate-mass (green; $0.5 < M/\Msun < 0.8$), and low-mass (blue; $M < 0.5\Msun$) stars in each episode. The left panel displays the data for GC1 (10 stripping episodes), while the right panel shows the data for GC2 (7 stripping episodes). The x-axis labels indicate the average time of each stripping episode.

For both clusters, the escaper population is dominated by low-mass stars. This is expected due to a few reasons. First, low-mass stars are much more numerous as a result of the IMF. Second, due to the mass segregation within the clusters, low-mass stars are, on average, on higher-energy orbits and are much more likely to be ejected through encounters. For GC1, there is a weak trend in which the fraction of low-mass escapers slightly decreases over time, consistent with the findings of \citet{balbinot2018}, though the effect is less pronounced. In contrast, GC2 does not exhibit a similar decreasing trend, showing instead a large amount of episode-to-episode scatter in the fraction of ejected low-mass stars.

We note that \citet{balbinot2018} compares two clusters on an orbit similar to that of Palomar 5, where the first cluster has a realistic present-day mass of $\sim4500\Msun$, and the second is three times more massive. In contrast, the present-day masses of our GC1 and GC2 clusters are only $\sim1.4$–$1.5$ times their ``initial masses'' at the time we begin tracking ejected stars (i.e., around 2 Gyr ago)—significantly less than the mass ratio in their comparison. Moreover, their lower-mass cluster is much closer to dissolution at $M_0 = 4500\Msun$, whereas both of our clusters have $M_0 > 10^4\Msun$. Finally, stellar mass-dependent trends in the escaper population may also depend on orbital properties, as we observe greater episode-to-episode scatter in GC2, which is on a more eccentric orbit. A systematic study involving a large number of clusters spanning a wide range of mass, orbital parameters, and mass-loss histories would help clarify the role of these factors in shaping the mass-dependent trends in the escapers.

\subsection{Detectability}
\begin{figure*}
\plotone{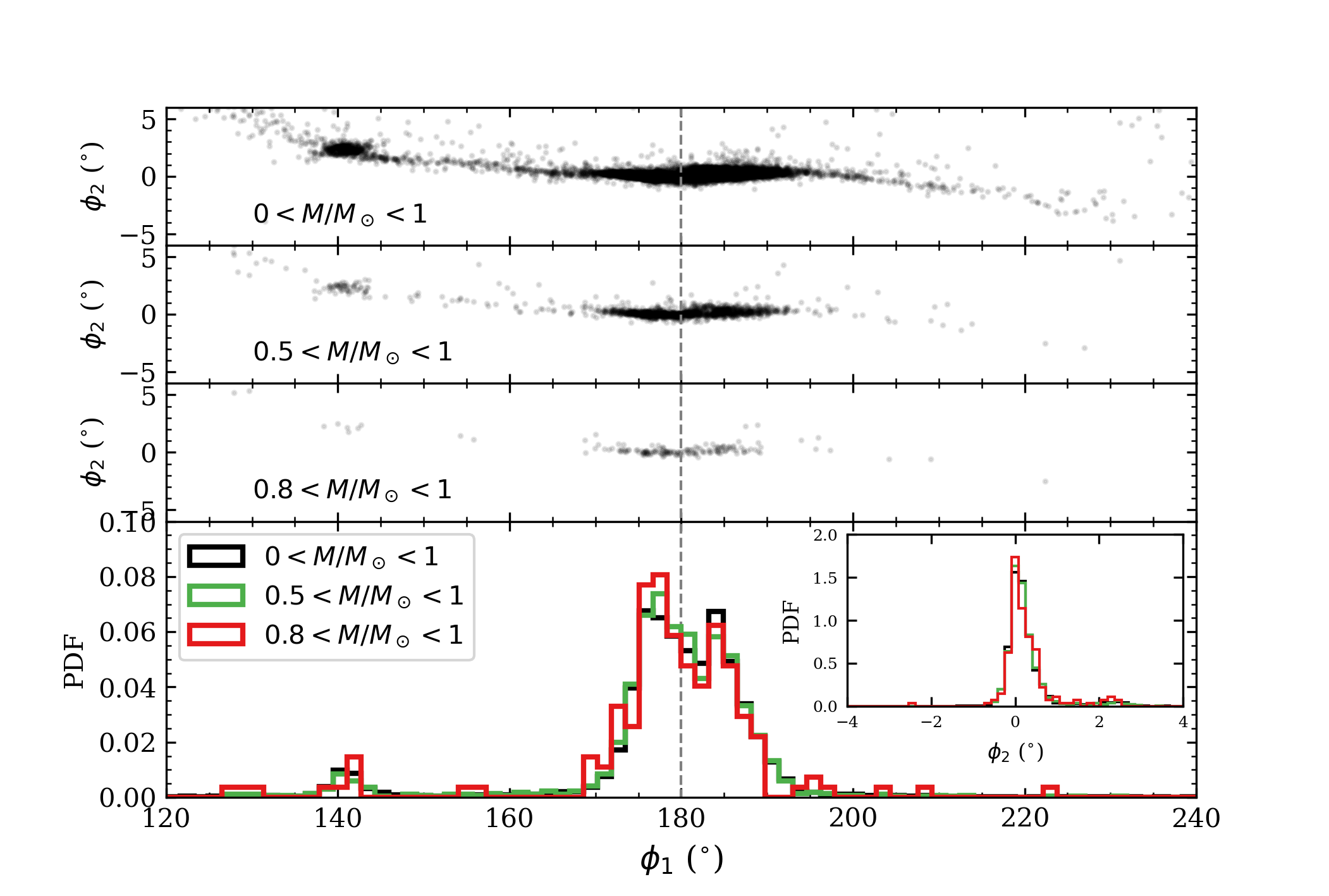}
\caption{The detectability of the GC1 stream as viewed in different stellar mass bins. The top three panels show stream stars with mass $0 < M/\Msun < 1$, $0.5 < M/\Msun < 1$, and $0.8 < M/\Msun < 1$, respectively. The bottom panel shows the normalized histograms of the stream stars in both $\phi_1$ and $\phi_2$. \label{fig:GC1_mass_seg}}
\end{figure*}

In Figure \ref{fig:GC1_mass_seg}, we show the GC1 stream (not including the stars that were recaptured) in its stream-aligned coordinates at the present day, separated by different stellar mass ranges. The top three panels display stream stars with $0 < M/\Msun < 1$, $0.5 < M/\Msun < 1$, and $0.8 < M/\Msun < 1$, respectively. These bins are chosen as proxies for stream detectability. Since stellar mass can be directly mapped to magnitude using an isochrone table---given an assumed stellar age and metallicity---these mass cutoffs can be inferred as different magnitude detection limits.

As the lower-mass cutoff increases, the number of detectable stream stars decreases significantly. At first glance, the stream appears shorter and narrower when only high-mass stars are detected. However, this is primarily due to the sheer number of low-mass stars, which results in a much higher number density of stream stars per unit area.

In the bottom panel, we show normalized histograms (PDFs) of the stream stars in $\phi_1$ (main plot) and $\phi_2$ (inset). The three different mass bins—$0 < M/\Msun < 1$ (black), $0.5 < M/\Msun < 1$ (green), and $0.8 < M/\Msun < 1$ (red)—are plotted for comparison. Interestingly, the PDFs in both $\phi_1$ and $\phi_2$ are nearly identical across all mass bins. We also perform a Kolmogorov–Smirnov test, which confirms that there is no statistically significant difference among them.

This finding suggests that the GC1 stream would appear to have the same width and length regardless of survey depth, assuming perfect stream membership selection. However, certain stream properties, such as velocity dispersion, might appear inflated in shallower surveys due to the lower number density of high-mass stars. Of course, real observations are more complex: factors such as stream star number density, contamination from background stars, dust extinction, and photometric uncertainties all affect the detectability of a given stream. We plan to generate detailed mock observations of GC streams under realistic conditions in future work.

%\FIXME{Maybe add something more quantitative here. For example, report the stream width (in degrees) for different mass bins and show that they are similar.}

\subsection{Black hole population in GC1 and GC2}

Thanks to our collisional GC model, we can also study the black hole (BH) population in GC1 and GC2. This is because the cluster's BH ejection timescale dictates its core density profile, which in turn can affect cluster escape and translate to observables along the streams \citep{roberts2025}. Stellar-mass BHs, which are more massive than the majority of stars in the GC, sink to the cluster's center after their formation, where they are likely to be ejected through strong encounters. The onset of the cluster's core collapse phase usually coincides with the time when all the BHs are ejected from the cluster.

More recently, a $33\Msun$ BH (\textit{Gaia} BH3) has been detected in the \textit{Gaia} DR4 pre-release data \citep{gaia-bh3-2024} and it has been found to be part of the ED-2 stream \citep{balbinot2024} that is currently passing through the solar neighborhood \citep{balbinot2023}. This is particularly exciting because it implies that black holes at these masses could form through interactions in dense star clusters.

The detection of a BH along the stream is unlikely in our model based on the two example clusters. Both GC1 and GC2 ejected all of their BHs within the first $\sim3$ and $\sim1$ Gyr of their formation, respectively. This occurred roughly $\sim8$ Gyr ago for both clusters, since GC1 is $\sim2$ Gyr older than GC2 in age. This ejection timeline is significantly earlier than the $\sim2$ Gyr threshold at which we begin orbit integration for the stellar escapers. Since the progenitor's orbit can change significantly even within the last 2 Gyr, it is unlikely that BHs ejected over 8 Gyr ago would still be located near the stream track traced by recent escapers or have dynamical properties consistent with the present-day progenitor. However, we note that our example clusters are not representative of the Milky Way globular cluster population (see Section \ref{subsec:limitations}). We will explore a larger sample of clusters spanning a wide range of ages, metallicities, and orbits in future work.

\subsection{Limitations}\label{subsec:limitations}
Here, we discuss the limitations and certain aspects that can be improved within \cosmogems.

\textit{The BFE frame rate:} In this work, we only consider stream stars that have escaped within the last 2 Gyr, even though the clusters have been losing mass for a much longer time. This limitation arises from our ability to accurately reconstruct stellar orbits using our set of BFE potentials, as the discrepancies between the true and reconstructed orbits for GC1 and GC2 become increasingly significant over time (Figure \ref{fig:GC12_orbits_true_vs_int}). We find that this is likely due to the frame rate of our BFE. The \textsf{m12i} simulation contains approximately 100 snapshots over the last 2 Gyr, corresponding to a frame rate of about 20 Myr between snapshots. This is roughly 10\% of the orbital timescale for GC1 ($\sim200$ Myr per orbit), which has the shorter orbital timescale of the two clusters.

We tested the convergence of the GC1 and GC2 orbit reconstructions by integrating the progenitors' orbits using various maximum BFE pole orders ($\ell_{\rm max}$) and timestep sizes. We find that the orbits generally converge for $\ell_{\rm max} \geq 4$ and timestep sizes of $\sim0.5$ Myr. We also experimented with higher-order integrators, such as a 4th-order Runge–Kutta method, but found no significant improvements, consistent with findings in \citet{arora2024}. However, re-running the \textsf{m12i} simulation with a higher snapshot frequency would require approximately 1 million CPU-hours \citep{wetzel2016}. We also note that we might be able to integrate some clusters beyond 2 Gyr with the current frame rate, especially larger halo orbits with longer orbital timescales.

\textit{The cluster evolution:} \CMC is much faster than traditional $N$-body codes because the calculations carry, at worst, an $\mathcal{O}(N\log N)$ complexity. Gravothermal effects are handled using a Monte Carlo method \citep{henon1971}, where the particles are sorted radially and two-body scattering is applied to neighboring particles. This adequately approximates the effect of many weak encounters. However, \CMC assumes that the cluster is spherically symmetric. This assumption might break down during tidal-debris production.  New Monte Carlo methods that break spherical symmetry are currently under development \citep[Tep et al., Cook at al., in prep][]{}.

%This motivates the development of the forthcoming {\sf KRIOS} code \citep[][Cook et al., in prep]{tep2025}. {\sf KRIOS} uses a self-consistent field \citep[SCF, e.g.][]{hernquist1992} to handle the collisionless dynamics for arbitrary progenitor configurations and supports several neighbor-finding routines such that H\'{e}non-style two-body relaxation can be applied.

%Moreover, our clusters were evolved using the tidal tensors and progenitor's orbit extracted directly from the tracer particle in the \textsf{m12i} simulation snapshot data. This is slightly 

\textit{The progenitor's self-gravity:} We model the progenitor's self-gravity using a Plummer model, which is generally acceptable due to the assumed spherical symmetry of the \CMC. However, a single Plummer model may not accurately capture the progenitor’s density profile at all evolutionary stages. In particular, \citet{weatherford2024} finds that a three-component Plummer model provides a much better fit to the steep central density region of a core-collapsed cluster. This distinction is especially important for clusters on circular orbits, which were the focus of \citet{weatherford2024}, since most stellar escapers in that regime originate from within the cluster's scale radius.
In contrast, both GC1 and GC2 are on eccentric orbits, where stars escaping from within the scale radius account for only $\sim10\%$ of the total stream population. When we remove these stars from our analysis, we find that the overall stream properties change only minimally.

%Alternatively, the cluster potential can be represented as a sum of basis functions \citep[e.g.,][]{zhao1996}. In \cite{tep2025}, the power law index $\alpha$ and scale length $b$ associated with the Zhao basis are fit to the stars' spatial distribution such that the optimal radial functions can be used to model the cluster mass density and potential scalar fields. Setting $\ell_{\rm max}=0$ enforces spherical symmetry and the zeroth-order basis function would likely be suitable in contexts similar to the analyses presented in this paper.

\textit{Comparison with MW GCs:} Our example clusters and the entire cluster population produced in GBoF1 and GBoF2 are not representative of the real MW GCs. This is because of the differences in the stellar population and the formation histories of \textsf{m12i} and the MW. Specifically, the model we used does not produce the old, metal-poor GC population likely because the star-formation rate in \textsf{m12i} is much lower than the MW at early times and is much higher than the MW at the present day (See Figure 13 in GBoF1 for the age-metallicity relation of these clusters). In future work, we will explore alternative GC formation models \citep[e.g.,][]{chen2022, chen2024} as well as simulations with early star formation, more similar to the MW.

%\FIXME{breifly summarize what Mike and Carl stated for the discrepancies, i.e., ages, metallicity etc.}

\section{Conclusions} \label{sec:conclusions}
We present a star-by-star cosmological GC stream model, \cosmogems, that self-consistently bridges small-scale cluster physics with large-scale Galactic dynamics. Specifically, the cluster formation model analytically maps the population of giant molecular clouds formed in the FIRE \textsf{m12i} simulation to cluster properties. Each cluster is then evolved using the collisional code \CMC to obtain a list of escaped stars, which are subsequently orbit-integrated in the combined cluster and host potential from the moment they meet the escape criteria to the present day.

We provide 2 example GC streams that we produce with this pipeline: GC1 and GC2. Both clusters are in relatively eccentric orbits and have the present day mass $>10^4\Msun$. Our main findings are:
\begin{enumerate}
    \item Due to their eccentric orbits, the mass loss of GC1 and GC2 is driven by episodic stripping near pericenter (Figure \ref{fig:GC12_prop}). This contrasts with nearly circular orbits, where mass loss is more uniform and primarily driven by cluster evaporation.
    \item The GC1 stream, on a less eccentric orbit, forms a relatively long and thin structure at the present day, with a velocity dispersion consistent with measurements from MW GC streams (Figure \ref{fig:dispersion}, bottom). Interestingly, it develops a local clump around $\phi_{1} = 140^\circ$ (Figure \ref{fig:dispersion}, top and middle). The origin of this clump is likely due to an interaction with the Galactic disk, as the GC1’s orbital plane lies close to the disk. This stream also exhibits an orbit-dependent misalignment in the stream track, which is most prominent near the progenitor’s apocenter (Figure \ref{fig:div_stream_track}). This mismatch arises from changes in the progenitor’s orbit over time (e.g., orbital plane, $r_{peri}$, $r_{apo}$). Stars that escape the cluster more recently tend to closely follow the progenitor's instantaneous orbit, which differs significantly from the orbits of stars that escaped earlier. The misalignment is less prominent near pericenter, as the stream is stretched due to the conservation of phase-space volume.
    \item The GC2 stream, on a more eccentric orbit, forms a multi-component structure consisting of a thin, stream-like segment and a diffuse, shell-like segment (Figure \ref{fig:GC12_streams}, bottom row). This multi-component feature has also been observed in MW streams, such as Jhelum \citep{bonaca2019}. In the case of the GC2 stream, we show that this feature naturally arises from stream stars spanning a wide range of orbital phases --- the thin component consists of stars near their fast-moving pericenter, while the diffuse component consists of stars near their slow-moving apocenter (Figure \ref{fig:GC2_orbital_phase}).
    \item Both the GC1 and GC2 streams exhibit little to no stellar-mass-dependent trends. For GC1, the fraction of ejected low-mass stars ($<0.5\Msun$) from each stripping episode decreases weakly over time (Figure \ref{fig:GC12_stripping_bar_chart}). This is not the case for GC2, where larger episode-to-episode variations may obscure any underlying trend.
\end{enumerate}

These interesting stream features arise from the interplay between cluster properties, their orbits, the Galactic model, and substructures within the Galaxy (e.g., the disk) --- all of which evolve over time in a non-trivial fashion. By using cosmological simulations, we are able to capture all of these time-dependent effects. However, unlike idealized simulations, it is much more challenging to isolate and study the impact of a single physical process on a stream in a cosmological setting. Therefore, for analysis at the level of individual streams, cosmological simulations work in complementarity with idealized simulations; that is, we can identify interesting features in cosmological streams, and then attempt to reproduce and test our hypotheses using idealized stream simulations.

More importantly, the streams we produced serve as an ideal dataset for testing automated tools commonly used to identify streams. Many of these tools assume static orbital properties for stream stars, which this work shows can vary significantly over time. For example, \textsf{STREAMFINDER} \citep{malhan2018} assumes that stream stars delineate the progenitor's orbit \citep{dehnen2004}, while the pole count map method \citep{mateu2017} assumes that stream stars lie along a great circle and exhibit clustering in their orbital poles. \textsf{Hough Stream Spotter} \citep{pearson2022} only searchers for 2D-projected linear features. As a result, machine-learning-based stream search techniques, such as \textsf{VIA MACHINAE} \citep{shih2022,shih2024}, that are agnostic to the underlying physics maybe able to outperform standard stream searches in some scenarios.

We note that the results presented in this work pertain to two example clusters and are not representative of the entire cosmological GC stream population. To determine whether these features are common among cosmological streams, a large sample of clusters spanning a wide range of parameter space is required. In future work, we will carry out a population-level analysis of the full GC stream population in the FIRE \textsf{m12i} simulation.

A clear advantage of cosmological simulations over idealized simulations lies in their capacity for population-level analysis. With cosmological models, \cosmogems complement recent studies \citep[e.g.,][]{chen2024,pearson2024} by incorporating time-evolving Galactic halo as well as a more realistic stellar escape scheme. We can begin to make predictions and address questions such as: How many GC streams exist around MW-like galaxies, and how many do we expect to observe in upcoming surveys? With thousands of galactic and extragalactic streams expected to be detected in the near future, this paper provides groundwork for understanding and modeling the cosmological GC stream population in the era of next-generation surveys such as Euclid, LSST, and Roman.

\begin{acknowledgments}
This work was made possible by NASA grant 22-ROMAN22-0055, and supported by NASA grant 22-ROMAN22-0013. %The second one is RINGS that Robyn, Tjitske, and Sarah are also a part of. Not sure we should include that one.
%Below added by SP
This work was supported by a research grant (VIL53081) from VILLUM FONDEN. Funded/Co-funded by the European Union (ERC, BeyondSTREAMS, 101115754). Views and opinions expressed are however those of the author(s) only and do not necessarily reflect those of the European Union or the European Research Council. Neither the European Union nor the granting authority can be held responsible for them.
%Below added by BC
This work was supported
by NSF Grant AST-2310362. NP is grateful for support from the Carnegie--Caltech Joint Postdoctoral Fellowship.
CLR acknowledges support from
the Alfred P. Sloan Foundation and the David and Lucile
Packard Foundation.  
BTC was partially funded by the North Carolina Space Grant's Graduate Research Fellowship.
%Below added by Tjitske
TS gratefully acknowledges the support of the NSF-Simons AI-Institute for the Sky (SkAI) via grants NSF AST-2421845 and Simons Foundation MPS-AI-00010513.
PFH was supported by a Simons Investigator Grant.
%thanks Mike and other people
This project was developed in part at the Streams24 meeting hosted at Durham University. We also thank Mike Grudi\'{c}, Nora Shipp, and Peter Ferguson for helpful discussion and input.
\end{acknowledgments}

%\facility{}
%\software{}

\bibliography{bibliography}{}
%\bibliography{main}{}
\bibliographystyle{aasjournal}

\end{document}